# Multi-Location Program Repair Strategies Learned from Past Successful Experience

Shangwen Wang[1], Xiaoguang Mao[1], Nan Niu[2], Xin Yi[1], and Anbang Guo[1]
[1]National University of Defense Technology, Changsha, China
[2]University of Cincinnati, Cincinnati, USA
{wangshangwen13, xgmao, yixin09, guoanbang12}@nudt.edu.cn, niunn@ucmail.uc.edu

*Abstract*—Automated program repair (APR) has great potential to reduce the effort and time-consumption in software maintenance and becomes a hot topic in software engineering recently with many approaches being proposed. Multi-location program repair has always been a challenge in this field since its complexity in logic and structure. While some approaches do not claim to have the features for solving multi-location bugs, they generate correct patches for these defects in practice. In this paper, we first make an observation on multi-location bugs in Defects4J and divide them into two categories (i.e., *similar* and *relevant multi-location bugs*) based on the repair actions in their patches. We then summarize the situation of multi-location bugs in Defects4J fixed by current tools. We analyze the twenty-two patches generated by current tools and propose two feasible strategies for fixing multi-location bugs, illustrating them through two detailed case studies. At last, the experimental results prove the feasibility of our methods with the repair of two bugs that have never been fixed before. By learning from successful experience in the past, this paper points out possible ways ahead for multi-location program repair.

*Keywords—automated program repair; multi-location bugs; case studies*

## I. INTRODUCTION

Over the years, researchers develop various *Automated Program Repair* (APR) techniques aiming at reducing the onerous burden of fixing bugs. Generally, these automated repair tools can be classified into two categories, i.e., search-based methodology (e.g., GenProg [1] and RSRepair [2]) and semantics-based methodology (e.g., S3 [3] and Angelix [4]). Search-based repair method (also known as generate-and-validate methodology) generate patch candidates by searching within a predefined fault space determined by Fault Localization (FL) techniques and then validate these candidates against the provided test-suite. Semantics-based repair methodology, on the contrary, utilizes semantic information generated by symbolic execution and constraint solving to synthesize patches. These state-of-the-art APR tools make great achievements on single-edit program repair.

Multi-location program repair, which refers to fixing multi-location bugs whose human-written patches contain multiple chunks [4, 5], has become a challenge since the rise of APR due to its complexity. Some recent empirical studies have shown the importance of multi-location repair: The study by Sobreira et al. [5] shows that more than 60% of the bugs in Defects4J [6], a well-known dataset containing 395 real bugs collected from six open-source Java projects, need fixing in multiple points; the study by Soto et al. [7] reports the median number of file changed in bug-fix commits in a large dataset from GitHub is 2; and Zhong and Su [8] draw the conclusion that programmers make at least two repair actions in total to fix more than 70% of bugs. So far, only Angelix and S3 have been reported to possess special features designed for multi-location bugs by capturing the dependence among multiple program locations. However, as the authors stated in [3], semantic-based repair exclusively modifies expressions in conditions or on the right-hand side of assignments, leading to its not so satisfactory performance. Thus, it is an emerging trend for solving multi-location bugs.

While some tools do not claim that they have the abilities to fix multi-location bugs, they generate correct patches for these bugs when being evaluated. For example, the patch in Fig. 1 changes two statements at different places into a same oracle-throwing statement and it is a typical multi-location defect[1]. Recently, a tool named ACS [9] has reported to fix this bug successfully. This phenomenon motivates our study. In this paper, we analyze why these patches are generated and how they fix the bugs, aiming to provide practical guidance for future research by learning the experience. We first conduct an empirical study on the multi-location bugs in Defects4J database and classify them into two categories according to repair actions in their patches. The benchmark Defects4J is chosen as our study subject since it is a wildly-used one for APR techniques on Java [9-18]. We then investigate the statistics of multi-location bugs from Defects4J that are successfully fixed by the current tools. We analyze the twenty-two patches for multi-location bugs generated by current tools and propose two strategies for solving this kind of bugs, illustrating them through two detailed case studies. The experimental results show the practicalities of our strategies with the repair of two out of eight multi-location bugs in Defects4J that have never been fixed before. Our methods are successful by micro-adjustment of our strategies with existing tools. To sum up, our contributions are:

- a posteriori classification of multi-location bugs in Defects4J based on the repair actions in their patches;
- a comprehensive summary of the situation of multi-location bugs fixed by current APR tools;
- two practical strategies for fixing multi-location bugs and two instances which are successfully fixed by our methods.

In the rest of this paper, we introduces our classification for Defects4J multi-location bugs in Section II. We investigate and summary multi-location bugs which have been fixed in Section III and provide our methods learned from the successful experience in Section IV. Section V demonstrates our experimental results. We discuss the paper in Section VI and make conclusion in Section VII.

---
[1] In this paper, we use the nouns "defect" and "bug" interchangeably.

## II. EMPIRICAL OBSERVATION

This section introduces a classification of multi-location bugs in Defects4J based on the repair actions in their patches. In this section and the rest of this paper, to refer to Defects4J bugs, we use a simple notation with project name followed by bug id, e.g., Math#5. Two patches for multi-location bugs in Defects4J, Lang#35 and Chart#5, are shown in Fig. 1 and Fig. 2, respectively. We divide multi-location bugs in this dataset into two categories by analyzing the repair actions in the two instances.

Note that in this paper, a line of code starting with "+" denotes a newly added line and lines starting with "-" denote lines to be deleted by developers.

In Fig. 1, developer changes two statements at different places into a same oracle-throwing statement. Modifications at each edit point share similar actions in syntax and thus this bug is classified into syntax similar multi-location bugs category (*similar multi-location bugs* for short). In this category, the similar modification we talk about may spread over both a single line of statement (Time#3) and a chunk of codes (Chart#14). Such cases are abundant in Defects4J, such as Time#3 where several similar *if* conditional statements are added, Math#49 where an object instantiation is modified in many functions from a same class, Closure#4 where a conditional expression is modified similarly at two places, etc. Among the 244 multi-location patches in our dataset, this category has 70 instances, occupying 28.69% of the total amount. A small part of these cases (23/70) share exactly the same operations at each edit point, such as Chart#14, adding the same conditional block in four edit points.

The fixing shown in Fig. 2 is different. It involves the addition of an *if* conditional statement and corresponding operations from lines 544-547 and the modification of the content of an *if* conditional statement in line 552. These modifications are compact and have great logical correlation in the program structure and we name this type semantic relevant multi-location bugs (*relevant multi-location bugs* for short). The criterion is from Abstract Syntax Tree (AST) level: *if the node under one modification appears in other modified places in this patch or the modified places are sub-nodes of a common node*, then the bug belongs to this type. Another example is the patch of Mockito#2 where developer first uses *Method Definition Addition* repair action from lines 30-34[2] and then operates *Method Call Addition* in line 11. This type is more popular in our dataset, holding 67.62% (165/244) of the total amount. This result is consistent with our perception that modifications performed at multiple locations aiming at solving a bug should have logical correlations, in most cases.

A small proportion of the dataset which contains 9 patches like Time#2 and Mockito#11 shows differences from the above two conditions: modifications at different places in these patches are neither similar nor logical related. We have not proposed any method for this kind of situation in this paper due

---

[2] Due to space limitation, some code snippets are not shown. Please check them in https://github.com/program-repair/defects4j-dissection.

Fig. 1. The patch of Lang#35

Fig. 2. The patch of Chart#5

to its peculiarity.

Note that the classification is based on the features of modifications at different places and it can provide guidance for repairing. *Similar* type bugs have no or weak logical correlation at each location and thus we may fix these places one by one, however, *relevant* type bugs possess strong logical correlations at each location and it may affect other places when operating in one place. That is why *relevant* type bugs are more difficult to repair and it is proved through the results which we will show in the next section: more *similar multi-location bugs* have been fixed than *relevant multi-location bugs*. Also note that when counting the number of each category, we use a *relevant first* strategy which means if a patch contains both similar and relevant edits, it belongs to the latter. For example, in Math#74, two similar loop chunks are added but there is another modification about the loops, making this patch belong to relevant bug. The reason for this strategy is that if both kinds of operations are needed for fixing a bug, then the difficulty degree is near to repairing a *relevant* bug.

## III. SITUATION STATISTICS

In this section, we investigate how the contemporary APR tools would handle the multi-location bugs. We select eleven tools which have been evaluated on Defects4J dataset: ProbabilisticModel (*PM*) [10], SimFix (*S*) [11], jGenProg (*jGP*) [12], jKali (*jK*) [12], Nopol (*N*) [13], ACS (*A*), ssFix (*ssF*) [14], JAID (*J*) [15], HDRepair (*HDR*) [16], SketchFix (*SF*) [17], and AVATAR (*AV*) [18]. There are some other tools aiming at repairing Java bugs such as JFix [19] and NPEFix [20]. We do not discuss them in this section since they have not been evaluated in Defects4J. We also overlook two latest tools (SOFix [21] and CapGen [22]) because both techniques are only designed for programs with a fault on a single point at this time as their authors clarify in the research papers. Hence, it is meaningless to include them in this investigation. Note that we adopt the experimental results for jGenProg, jKali, and Nopol

TABLE I. STATISTICAL RESULTS

| Bug ID | T | PM | S | jGP | jK | N | A | ssF | J | HDR | SF | AV |
|---|---|---|---|---|---|---|---|---|---|---|---|---|
| C5 | R | | | | | | ✓ | | | | | |
| C14 | S | | | | | | ✓ | | | | | |
| C19 | S | | | | | | ✓ | | | | | |
| CL115 | R | | | | | | | ✓ | | | | |
| L10 | R | | | | | | | | | ✓ | | |
| L27 | R | | ✓ | | | | | | | | | |
| L35 | S | | | | | | ✓ | | | | | |
| L41 | R | | ✓ | | | | | | | | | |
| L50 | S | | ✓ | | | | | | | | | |
| L60 | S | | ✓ | | | | | | | | | |
| M4 | S | | | | | | ✓ | | | | | |
| M22 | S | | | | | | | | | ✓ | | |
| M35 | S | | ✓ | | | | ✓ | | | | | |
| M61 | R | | | | | | ✓ | | | | | |
| M71 | S | | ✓ | | | | | | | | | |
| M79 | R | | ✓ | | | | | ✓ | | | | |
| M90 | R | | | | | | ✓ | | | | | |
| M93 | R | | | | | | ✓ | | | | | |
| M98 | S | | ✓ | | | | | | | | | |
| M99 | S | | | | | | ✓ | | | | | |
| *Similar* | 11 | - | 5 | - | - | - | 6 | - | - | 1 | - | - |
| *Relevant* | 9 | - | 3 | - | - | 1 | 3 | 2 | - | 1 | - | - |
| Total | 20 | - | 8 | - | - | 1 | 9 | 2 | - | 2 | - | - |

Column "Bug ID" uses a brief notation of the bug due to the space limitation. Please note that *C* denotes the project *Chart* and *CL* (in the fifth line) denotes the project *Closure*. Column "T" means the type of this bug and R refers to *relevant* type while S refers to *similar* type. "✓" denotes this bug is successfully fixed by the tool. "Similar" and "Relevant" denote the numbers of different types of bugs fixed by each tool and "Total" denotes the total number of bugs fixed by each tool. It is marked with "-" if the tool cannot fix any bug.

reported by Martinez et al. [12] and the results of other approaches come from the corresponding research papers. The results are illustrated in Table I where each tool is represented by its acronym.

Please note that the authors define a concept *Partially-fixing* in the study [18] which means patches that make the program pass part of the previously-failing test cases. It is reported that AVATAR partially fixed some multi-location bugs like Chart#14, however, in our investigation, we do not consider these partially-fixings as correct patches as they still do not pass all the test cases in the test suite.

Generally speaking, 22 valid patches are generated and 20 multi-location bugs are successfully fixed including 11 *similar* type and 9 *relevant* type, among which Math#35 and Math#79 are fixed by two tools. There are only five tools being able to fix these bugs (i.e., SimFix, Nopol, ACS, ssFix, and HDRepair) among which SimFix and ACS repair the most bugs (8 and 9, respectively). Nopol and ssFix repair 1 and 2 bugs respectively and they can only fix *relevant multi-location bugs* at this moment. HDRepair fixes one bug for each type.

## IV. LESSONS LEARNED AND STRATEGIES

In this section, we propose two strategies learned from successful experience provided by the twenty-two correct patches generated by current tools and illustrate them through two detailed case studies which analyze the patches generated for the two bugs shown in Section II.

### A. Case Study 1: Patch of Lang#35 Generated by ACS

We list the patch of Lang#35 generated by ACS in Fig. 3. Another modification chunk performed at line 3578 is the same as the code in the figure. ACS is especially designed for synthesizing conditional expressions containing two steps: variable selection and predicate selection. It uses a method named *Oracle-Throwing* to avoid the crash, thus, it can generate patch as shown. However, why it is able to generate two modification chunks still needs further explanation. Note that Lang#35 is a *similar multi-location bug* which means the two modification points have no or weak correlation in program structure. Thus, there may be multiple test cases aiming at testing different methods in the program and they all fail. We find that the test suite for this project contains two failing test cases and when executing, ACS uses a *fitness function* which enables it to continue fixing based on a partial-fixing program if the repair actions that have been performed reduce the number of failing test cases (Please note that ACS does not introduce this feature in its paper, we get this information after connecting with the authors). That is the key point for generating this patch. Fitness function is wildly used in APR techniques. Previous studies such as GenProg and HDRepair use fitness functions to guide the selection process of candidate patches while ACS exploits the deduction of failing test cases for solving buggy points one by one, bringing a new idea for *similar multi-location bugs*. The main challenge for applying this strategy is the precondition: the test suite must have enough failing test cases to expose the defects which means the test suite must be augmented. This insight is consistent with recent studies [23, 24] where the need of augmented test cases is proposed by the authors. Test case purification [25], which means recovering the execution of omitted assertions, has the ability to generate more practical test cases and enhance the performance of test suite. EvoSuite is a commonly used tool for automated test suite generation and empirically, it can increase code coverage up to 63% [26]. For programs whose test cases are written by developers, if we first add test cases generated by Evosuite into the test suites and conduct test case purification, leading to an augmented test suite, and then use this fitness function to repair, we may be able to solve more multi-location bugs. This strategy is suitable for our study subject, Defects4J, since all the projects in this benchmark are open source projects and the original test suites are manually created. Note that solving the defect in one location may increase the number of failing test cases coincidentally, thus the fitness function should possess resilience to prevent the correct patch from being screened out.

> **Strategy 1**: For *similar multi-location bugs*, use a suitable fitness function which are able to continue the repair process based on a partial fixing program combined with augmented test suites.

```
3300        //ACS's patch begin
3301   +    if (element == null){throw new IllegalArgumentException();}
3302        //ACS's patch end
3303        return newArray;
3304   }
```

Fig. 3.  Patch of Lang#35 generated by ACS

```
--- /tmp/chart_5_Nopol/source/org/jfree/data/xy/XYSeries.java
+++ /tmp/chart_5_Nopol/source/org/jfree/data/xy/XYSeries.java
@@ -562,3 +562,3 @@
           // append the value to the list...
-         if (this.autoSort) {
+         if (overwritten!=null) {
               this.data.add(-index - 1, new XYDataItem(x, y));
```

Fig. 4.  Patch of Chart#5 generated by Nopol

### B. Case Study 2: Patch of Chart#5 Generated by Nopol

The patch of Chart#5 generated by Nopol is shown in Fig. 4. Unlike the human-written patch shown in Fig. 2 modifying two code chunks, this patch only modifies a conditional statement to repair this bug. The modification point is at line 562, just under the buggy point. Nopol is a semantic-based program repair tool utilizing angelic values and a Satisfiability Modulo Theory (SMT) solver for synthesizing conditional expressions. The conditional expression it generates really avoids the error. The variable *overwritten* is defined with *null* in line 546 and its value can only be modified if the condition in line 548 is met. When the condition in line 548 is not met, *overwritten* keeps the value *null* and the program goes to the conditional branch in line 562 where the condition is not satisfied, either, after being modified. Then the program skips this conditional branch and goes to line 566 directly and thus the wrong expression in line 548 does not cause the error in line 564 which means the error is avoided. This strategy, generating guard preconditions to avoid the potential faults, is to some extent like **fault tolerance** technique [27]. We further study why the modification at line 562 makes sense.

Recently, **Error Propagation Chain** (EPC), which refers to a sequence of statements between program defect and program failure statement, is proposed by Guo et al. [28] to improve the efficiency of fault localization. We check their experimental results for Chart#5 and find that line 562 is in this chain. That indicates a new direction for fixing *relevant multi-location bugs*: since modifications at each edit point possess correlation in logic and it is hard for current technologies to fix at each point, we can find out the closest intersection to the buggy points in the EPCs and utilize SMT solver to find a patch which adds a guard precondition to avoid the error. It is possible to generate a patch as long as an intersection can be found no matter how many buggy points the program possesses. For applying this strategy, we need to select out the top-k suspicious statements and calculate the intersections in their EPCs. If the intersections have already been included in the suspicious statements list, we do not perform any operation; otherwise, we add these intersections into the suspicious statements list. Then we can rerun the APR tool for generating a patch. In our experiments, we empirically set k to 100. Note that two situations of *relevant multi-location bugs* have been introduced in Section II. If two AST nodes have dependency relations, one statement will appear in another's EPC; if two nodes are both under a common node, then there will be an intersection in their EPCs. That is to say, for a relevant multi-location bug, there is at least one intersection in the EPCs of buggy points.

> **Strategy 2**: For *relevant multi-location bugs*, find out the intersections of the EPCs of buggy points and search for modifications at these points.

## V. EXPERIMENTAL RESULTS

In this section, we introduce our experiment design and results.

### A. Experiment Setup

We chose to randomly select out four *similar multi-location bugs* whose test cases are not yet capable for exposing all the defects and four *relevant multi-location bugs*, each of whose EPC is less than ten lines, from Defects4J to conduct our experiment. The selection is completely random without any bias and all the selected samples have never been repaired before. The reason for this evaluation subject is that the state-of-the-art APR tools still possess low recall on Defects4J benchmark (the highest, SimFix, is lower than 10%). Thus, even if we consider all the suitable multi-location bugs, it is still unrealistic to expect plenty successful cases. However, if there is a success in our sampled subject, the feasibility of our strategies is proved.

For *similar multi-location bugs*, we used SimFix to conduct the experiments because 1) ACS can no longer execute on new bugs now as it announces in its homepage [3], 2) SimFix possesses the same fitness function as ACS which enables it to perform multi-location fixing (probably because they are developed by the same group) and 3) SimFix itself integrates the test case purification technique which can extremely simplify our experiments. This part was performed on a 64-bit Linux virtual machine with Ubuntu 15.10 operating system and 2GB RAM. For *relevant multi-location bugs*, we chose Nopol since it can synthesize *if condition statement*. This part was performed on a 64-bit Linux host with Ubuntu 14.04 operating system and 32 GB RAM.

Both experiments are two-phase: for *similar multi-location bugs*, we first added test cases generated by Evosuite into the test suite and then ran SimFix; for *relevant multi-location bugs*, we first calculated the intersections in the EPCs of their top-100 suspicious statements (we used the algorithm introduced in [28] to calculate EPC) and conducted corresponding operations as we have introduced in Section IV.B, then we exploited Nopol to generate patches.

The testing framework and FL ranking metric for both experiments are GZoltar[4] version 1.6.0 [29] and Ochiai. A recent study [30] shows that fault localization step may affect the evaluations of APR tools. Obeying their suggestions, we added all the lost location information when conducting experiments and thus made the experiment *Line_Assumption* which means the faulty code lines are known. Our intuition is that if the faulty points are not accurately located, it is less likely to repair multi-location bugs. A *Line_Assumption* experiment is free from the bias caused by FL step and can explicitly check the performance of the repair methodology.

---
[3] https://github.com/Adobee/ACS
[4] http://www.gzoltar.com/

TABLE II.  EXPERIMENTAL RESULTS

| Bug ID | SimFix + Strategy1 | | | | Nopol + Strategy2 | | | |
|---|---|---|---|---|---|---|---|---|
| | Math#46 | Math#49 | Lang#62 | Time#3 | Math#79 | Closure#8 | Closure#50 | Lang#22 |
| Execution time | 301min | 211min | 226min | 307min | 46s | 76s | 67s | 11s |
| Result | Timeout | Success | Timeout | Timeout | No Synthesis | Success | No Synthesis | No Angelic Value |

### B. Results and Analysis

The experimental results are shown in Table II. Generally speaking, two patches are generated where one is for *similar multi-location bug* and another one is for *relevant multi-location bug*. We manually examine the generated patches and consider a patch correct if it is the same or semantically equivalent to human-written one (this criterion is widely-used in recent studies [11, 18, 22]). Results show that both patches are correct. Note that our experiment is two-phase and the first phase is manually conducted, thus the execution time recorded in the table only refers to the time consumption of the second phase. Since SimFix is a search-based tool which means there will be a large amount of calculation and validation during the execution, its average execution time is much longer than that of Nopol.

Test case purification focuses on recovering the execution of omitted assertions and thus is sometimes useless for strengthening the test suite. For example, in Math#49, the test case *OpenMapRealVector* fails for its first function invocation which leads to an *InvocationTargetException* and thus the following function invocations cannot be executed which means the original test suite does not expose all the faulty points, being the reason for SimFix not fixing this bug. In our experiment, we added the test cases generated by Evosuite into test suite, successfully exposed the two defects, and at last fixed this bug. The generated patch is the same as the standard one provided by Defects4J. We performed the same operation to the other three *similar* type bugs but SimFix failed to generate patches for them mainly because two reasons. For Lang#62 and Time#3, the reason is SimFix finds for fixing ingredients in the original projects but there is no correct fixing ingredients in the source files. For example, Time#3 needs a mathematic symbol != but is does not exist in the whole project. This phenomenon indicates that we may combine source files with existing open source projects to enlarge the space for searching for fixing ingredients in the future. For Math#46, the reason is the code snippet is so large (10 lines) that it considers the donor with a return statement the same as human-written patch as not similar. However, after we adjusted the code snippet size to a finer-grained value (2 lines) and reran this bug, it still neglected the snippet which contains fix ingredient, which indicates that a more validate way for donor snippet identification should be developed. Both two findings are consistent with a recent study [31] which evaluates SimFix on Mockito project.

In Closure#8, the edit point in human-written patch (i.e., line 202 in class *CollapseVariableDeclarations*) is not included in the localization result generated by GZoltar, the FL technique used by Nopol. As the experiment is *Line_Assumption*, we calculated its EPCs and manually added these lines into the suspicious statement list. Finally, Nopol generated a patch under the class *Node* which is semantically equivalent to the human-written one. The other three *relevant* type bugs failed because of not finding angelic value at the interpoints (Lang#22) and not synthesizing a patch (Math#79 and Closure#50), corresponding to two of five limitations (*No angelic value found* and *Timeout in SMT*) the authors discussed in their paper, which means the repair ability of Nopol needs to be improved.

By making some micro-adjustments of our strategies with current tools, we fixed two multi-location bugs. The failed cases are due to the weaknesses of current tools according to our analysis, indicating the potential of our strategies to fix more bugs when combined with more powerful tools.

### VI. DISCUSSION

**Limitation:** We focus on the bugs in the Defects4J dataset, thus the generality of our methods to other multi-location bugs is unknown. Besides, we only select part of the multi-location bugs to conduct our experiment. This threat is limited given that our purpose is to prove the feasibility of our strategies. Since state-of-the-art techniques still have limitations, we do not expect our methods can solve plenty bugs at this time.

**Related Work:** Although APR is a hot topic in Software Engineering with a new APR system being proposed every couple of months, most state-of-the-art APR tools focus on single-location bugs, e.g., CapGen performs a single mutation to generate patches. GenProg can change multiple locations but the study [32] shows that the majority of cases are functionally equivalent to single line modification. Angelix and S3 design a feature for multi-location bugs by simplifying the angelic forest. The study [10] mines human patches to identify edits that commonly occur together in human-written patches to provide the first step for traversing the large search space for fixing multi-location bugs, providing another idea for solving this problem. Our study starts from the existing successful patches, concludes useful experience, and provides feasible strategies for future research.

**Future Work:** We divide multi-location bugs into two types and propose methods for each type. However, the type of a bug is decided through its edits, which is not available before the program repair process. In the future, we aim to develop a method to solve this problem by analyzing the context information of each suspicious point. Further, we will integrate this classification method with our strategies and design a tool for automated multi-location program repair.

## VII. Conclusion

In this paper, we divided multi-location bugs in Defects4J into two categories according to the repair actions in their patches, summarized the situation of these bugs fixed by current tools, and learned the successful experience as well as put forward two strategies for future research (one for each type). Guided by our strategies, we successfully fixed two multi-location bugs in Defects4J which have never been repaired before. To our best knowledge, we are the first to propose strategies by analyzing patches generated by current tools, bringing new idea for APR techniques as well as pointing out possible ways for multi-location program repair.

## Acknowledgment

Our work is supported by the National Natural Science Foundation of China (Grant No.61379054, 61672592).